# Self-assembly of complex salts of cationic surfactants and anionic-neutral block copolymers. Dispersions with liquid-crystalline internal structure

*Letícia Vitorazi[1], Jean-François Berret[2,]\* and Watson Loh[1,]\**

[1]Institute of Chemistry, Universidade Estadual de Campinas (UNICAMP), Caixa Postal 6154, Campinas, São Paulo, Brazil.
[2]Matière et Systèmes Complexes, UMR 7057 CNRS Université Denis Diderot Paris-VII, Bâtiment Condorcet, 10 rue Alice Domon et Léonie Duquet, 75205 Paris, France.

**Abstract**

We report the synthesis of complex salts made from the cationic surfactant dodecyltrimethyl-ammonium and diblock copolymers poly(acrylic acid)-block-poly(acrylamide) of different molecular weights. In water, the complex salts self-assemble into stable hierarchical aggregates with a dense core and a diffuse shell. In contrast to earlier reports, the surfactant/polymer aggregates exhibit a liquid crystalline structure of $Pm3n$ cubic symmetry. The crystal structure is analogous to that obtained with homopolymer. Size and aggregation numbers were estimated from a combination of light and small-angle x-ray scattering experiments. It is found that the size of the aggregates decreases with increasing diblock asymmetry. The complex salt methodology presents many advantages, among which to be insensitive to the preparation conditions and to the mixing pathway.

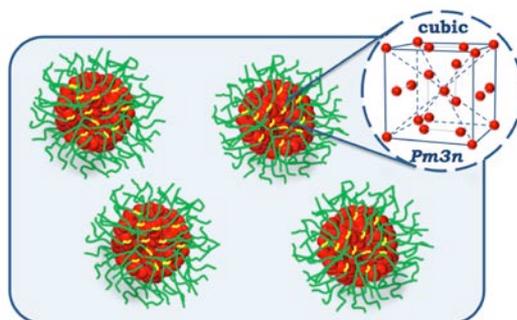

# I – Introduction

The self or co-assembly of macromolecules and colloids is a physicochemical process ubiquitous in nature.[1] It involves a range of non-covalent interactions (hydrogen bonding, electrostatic and hydrophobic interactions) which act to yield well-defined structures of tens-of-nanometers to a few microns size. With electrostatics, the assembly process is driven by the pairing of the positive and negative charges located at the surface of colloids or along macromolecules. The association between oppositely charged species is effective and leads to an associative phase separation.[2–4] The phase separation can be however prevented by modifying the formulation process[5] or by using specific polymer architectures.[6,7] It is the case when ion-containing polymers are replaced by hydrophilic charged-neutral block copolymers. Unlike amphiphilic copolymers, charged-neutral blocks are soluble in aqueous media, and co-assemble in the presence of oppositely charged species. These copolymers were discovered in the mid 1990's with the synthesis of polyion complex micelles made from block polypeptides with poly(ethylene glycol) segments.[8,9] These colloids form spontaneously by direct mixing and exhibit a core-shell structure. The growth is arrested at a size that is fixed by the dimension of the polymer. With charged-neutral block copolymers, electrostatic-based complexes were prepared using a wide range of systems, including polymers, colloids, biological molecules, or multivalent counterions, generating structures of great interest for applications in drug[9–14] or gene delivery.[15–17]

During the last years, a series of studies reported the formation of core-shell aggregates made from oppositely charged surfactants and charged-neutral block copolymers.[6,7,18–22] The ratio between the positive and negative charges present in the medium was found to have a strong impact on the assembly process.[19] Isothermal Titration Calorimetry performed on the poly(sodium acrylate)-block-poly(acrylamide)/dodecyltrimethylammonium bromide mixed system revealed that the electrostatic co-assembly depends on the mixing order, and that the reaction between the charges is endothermic.[21] The endothermic character of the reaction suggests that it is controlled by the entropy of the released counterions. On the same system, it was also found that the amount of polyelectrolytes needed for the complex formation exceeded the number necessary to compensate the net micellar charge, confirming the evidence of overcharging and non-stoichiometric complexes.[20,21]

Additional studies on polymer/surfactant systems[7,20,23,24] have shown that the micelles in the cores are in a disordered state, and display no long-range order. The absence of ordered structure was attributed to the fast kinetic process occurring during the mixing. Using poly(ethylene oxide-*b*-poly(methacrylic acid) and N-dodecylpyridinium chloride, Uchman and coworkers reported recently the formation of core-shell aggregates with liquid crystalline interior, a process that was achieved by fine-tuning the charge ratio between constituents.[22,25] A different kind of phase segregation was also investigated using alkyltrimethylammonium amphiphiles and diblock copolymers of poly(styrene)-*b*-poly(methacrylic acid) in the melt state. Various grafting densities of amphiphilic molecules, corresponding to different charge ratios were analyzed and different morphologies of cubic, hexagonal and lamellar symmetry were observed.[26]

An original strategy for the preparation of surfactant/polymer aggregates was presented some years ago.[2,27] By letting the cationic surfactant in its hydroxide form react with the acidic form of the polymer, and by removing the solvent, a pure component system devoid of counterions can be achieved. This complex salt allows the determination of true binary phase diagrams for a variety of mixtures. It also enabled the assessment of thermodynamic contributions of the surfactant and polyions, as well as on the effect of added organic solvents or electrolytes.[27–32] Inspired by this methodology, we re-examined the earlier investigations on the co-assembly between poly(acrylic acid)-*b*-poly(acrylamide) and dodecyltrimethylammonium bromide and compare the results with those obtained by other mixing pathways. The role of the asymmetry of the diblocks is also emphasized here by an appropriate choice of molecular weights. As a result, the size of the aggregates was found to decrease with increasing asymmetry. This new methodology also allows the preparation of core-shell aggregates with liquid crystalline interiors of various colloidal sizes.

## II – Experimental Methods

### II.1 – Materials

For the synthesis of polymers and complex salts, the following compounds were used: potassium ethyl xantogenate (purity 96%, Aldrich), methyl 2-bromopropionate (purity 98%, Aldrich), acrylic acid (stabilized, purity 99.5%, Acros Organics), acrylamide (purity ≥ 98%, Fluka), 2,2-azoisobutyronitrile (AIBN, purity ≥ 98%, Fluka), $CDCl_3$ (99.8% deuterium + 0.05 vol. % TMS, Cambridge Isotope Laboratories), dodecyltrimethylammonium bromide ($C_{12}TAB$, Aldrich). Acrylamide and AIBN were recrystalized using methanol as solvent. All other solvents and chemicals were used as received. In this work, poly(acrylic acid)-b-poly(acrylamide) block copolymers, abbreviated as $HPA_x$-b-$PAm_y$ were synthesized by controlled radical polymerization in solution using xanthate as chain transfer agent (S1). Their molecular weights were 7000 g mol$^{-1}$ for the HPA block and 6600 or 26000 g mol$^{-1}$ for the PAm blocks, yielding the developed formulae $HPA_{100}$-b-$PAm_{93}$ and $HPA_{100}$-b-$PAm_{366}$. These two polymers were compared to two other chains, $HPA_{70}$-b-$PAm_{422}$ and $HPA_{70}$-b-$PAm_{844}$, which were provided to us by Solvay.[7,23,33] Their molecular weights were 5000 g mol$^{-1}$ for the HPA block and 30000 or 60000 g mol$^{-1}$, respectively (Table 1). Throughout the paper, the polymers are characterized by their asymmetry ratio, which is calculated as the ratio between the degrees of polymerization of the charged and neutral blocks, HPA and PAm respectively. As a result, for the polymers mentioned previously, the asymmetry varies from 1.1 for $HPA_{100}$-b-$PAm_{93}$ to 0.08 for $HPA_{70}$-b-$PAm_{844}$ (Table 1).

| Polymers | $M_w$ (g mol$^{-1}$) | $M_w/M_n$ | Asymmetry | Ref. |
|---|---|---|---|---|
| $HPA_{100}$ | 7200 | 1.6 | ∞ | this work |
| $HPA_{100}$-b-$PAm_{93}$ | 14000 | 2.4 | 1.1 | this work |
| $HPA_{100}$-b-$PAm_{366}$ | 33000 | 3.2 | 0.27 | this work |
| $HPA_{70}$-b-$PAm_{422}$ | 35000 | 1.6 | 0.16 | 19 |
| $HPA_{70}$-b-$PAm_{844}$ | 65000 | 1.6 | 0.08 | 19 |

**Table 1.** Molecular characteristics of homo- and block polymers synthesized in this work.

### II.2 - Synthesis and characterization of block copolymer

The xanthate or RAFT agent, S-(2-methyl propionate) O-ethyl xantate, was prepared as previous described by Pound using methanol instead ethanol as solvent.[34] The product was characterized by $^1$H-NMR. $^1$H-NMR spectrum of the RAFT agent previously dissolved in deuterated chloroform was obtained at 25 °C on a 250 MHz Bruker spectrometer (S1). Trimethylsiloxane (TMS) was used as reference. $^1$H-NMR 250 MHz spectra showed the following features: 4.63 (m, 2H from C(S)OCH$_2$ moiety), 4.38 (m, 1H from CH moiety), 3.76 (s, 3H, from OCH$_3$), 1.58 (d, 3H from CH$_3$) and 1.42 (t, 3H from CH$_3$), where the abbreviations s, d, t and m denote singlet, doublet, triplet and multiplet, respectively.

*Synthesis and characterization of polymer and block copolymers:* The polymers were prepared by Reversible Addition-Fragmentation chain Transfer (RAFT) as previous described.[35] To prepare the HPA homopolymers, the acrylic acid monomers and S-(2-methylpropionate) O-ethyl xanthate were dissolved in 1,4-dioxane. AIBN in 1,4-dioxane was slowly added drop wise under argon atmosphere into the reaction mixture at 70 °C for 13 h. The homopolymer (macro chain RAFT) was isolated from the solution by precipitation using excess of hexane and recovered, as a powder by solvent removal under reduced pressure. The block copolymer was prepared dissolving the dried macro chain RAFT, acrylamide and AIBN in 1,4-dioxane. The reaction was carried out under argon atmosphere at 70 °C for approximately 18 h. The reaction mixture was equilibrated at room temperature and filtered. The crude product was thoroughly washed with 1,4-dioxane. In this way, the soluble unreacted HPA homopolymers still present in the synthesis bath could be safely removed and separated from the diblocks. Molecular weight distributions of the prepared homo- and block copolymers were determined by gel permeation chromatography (GPC) using a Viscotek GPCmax VE 2001, with double Viscotek detectors (refractive index – VE3580 and ultraviolet – 2500) and a set of three SB-806M HQ columns (S2). The columns were calibrated with PEO standards of narrow dispersity (from 630 to 584000 g mol$^{-1}$). The polymers were dissolved in aqueous solution containing NaSO$_4$ at 0.05 M and analyzed in a flow rate of 1mL min$^{-1}$ with an injected volume of 100 μL. The characteristics of homo- and block copolymers are summarized in Table 1.

*Preparation of the complex salts:* The acid form of HPA$_x$-*b*-PAm$_y$ in aqueous solution was titrated with the hydroxide form of cationic surfactant solution in water following the general procedure described earlier.[36] Aliquots of HPA$_x$-*b*-PAm$_y$ dissolved in water were titrated with surfactants using a standard glass electrode for the pH measurement. The equivalence point at pH 8.6 was found to be identical to that determined with the homopolymers.[36] For the preparation of the complex salt, the surfactant hydroxide solution was added drop wise to the HPA$_x$-*b*-PAm$_y$ solution under constant stirring until the equivalence point was reached. The mixture was left overnight at 4 °C. The complex salts, abbreviated as C$_{12}$TA(PA$_x$-*b*-PAm$_y$) were freeze-dried and the powders were kept in a desiccator. Complex salt solutions obtained by dissolving the C$_{12}$TA(PA$_x$-*b*-PAm$_y$) powder at the desired concentration.

**II.2 - Methods**

*Light scattering experiments:* The samples were evaluated by dynamic (DLS) and static (SLS) light scattering measurements. For measurements performed as a function of the scattering angle *θ*, experiments were conducted on an ALV-7004 spectrometer (ALV GmbH, Langen, Germany). C$_{12}$TA(PA$_x$-*b*-PAm$_y$) complex salts and deionized water were weighed and mixed directly in clean light scattering cells of borosilicate glass. The measurements were carried out with a 632.8 nm laser and *θ* was varied from 30 to 150° with an increment of 20°. For experiments aiming at testing the mixing pathway, samples were prepared at 1, 5 and 10 wt.%, and the 5 and 10 wt.% specimen were diluted to 1 wt.% for the measurements. Other samples were prepared by mixing the two components using a Thurrax T 10 basic dispersion equipment from IKA® at speed 3 (6500 rpm) and 5 (16500 rpm) for 1 min. These samples were evaluated at 25 °C by DLS using a Malvern Nano Zetasizer instrument with 632.8 nm laser and detector positioned at 173°. The light intensity of autocorrelation function was analyzed by the cumulant method which provided the Z-average radius and dispersity indexes. For dynamic measurements, the second-order autocorrelation function of the scattered light $g^{(2)}(t)$ was measured and analyzed using the constrained regularization REPES algorithm (Regularized Positive Exponencial Sum). The apparent diffusion coefficient $D_0$ of diluted solutions was determined according to $\Gamma(q) = D_0 q^2$, where $\Gamma(q)$ is the decay rate of the correlation function at the scattering wave-vector $q = \frac{4\pi n}{\lambda} \sin \frac{\theta}{2}$. In the previous expression, $n$ denotes the refractive index of water and $\lambda$ is the laser wavelength. The relaxation time distributions $\tau A(\tau)$ were obtained by inverse Laplace transformation ($\tau = 1/\Gamma$). From the $D_0$-values, the hydrodynamic radius $R_H$ of aggregates was determined using the Stokes-Einstein relationship for translational diffusion:

$$R_H = \frac{k_B T}{6\pi \eta D_0} \qquad (1)$$

where $k_B$ is the Boltzmann constant, $\eta$ is the viscosity of water and $T$ is the absolute temperature. For the static light scattering data, the Berry representation was used to display the wave-vector dependence of the Rayleigh ratio $R(q,c)$:[37]

$$\sqrt{\frac{Kc}{R(q,c)}} = \frac{1}{\sqrt{M_W^{App}}} \left(1 + \frac{q^2 R_G^2}{6}\right) \qquad (2)$$

where $c$ is the concentation of scatterers, $K = 4\pi^2 n^2 (dn/dc)^2 / N_A \lambda^4$ the scattering contrast ($N_A$ is the Avogadro number), $M_W^{App}$ the apparent molecular weight of the aggregates and $R_G$ their gyration radius. Eq. 2 was found to be more accurate for colloids or macromolecules which radius of gyration was of the order and larger than 100 nm.[37] For the calculations of the parameter $K$, the refractive index increments $dn/dc$ for the mixed aggregates was assumed to be a linear function[38] of the increments of the different components (S3). For HPA, PAm and $C_{12}$TAB, we used $dn/dc$ = 0.149, 0.185 and 0.153 mL g$^{-1}$, respectively.[39,40,33]

*Small-angle X-ray scattering (SAXS):* Appropriate amounts of the complex salt and Milli-Q water were weighed into glass tubes that were flame sealed. The mixing was performed in a centrifuge at 4000 rpm and 25 °C (various cycles of up-side-down inversion) during several hours to achieve the homogenization of the sample. These samples were analyzed by SAXS using the D11A-SAXS1 beamline of the Brazilian Synchrotron Light Laboratory (CNPEM, Campinas, Brazil). The samples were placed into a sealed sample holder closed by two mica windows and the scattering patterns were recorded under vacuum. The scattered intensity $I(q)$ was obtained with scattering vector $q$ ranging from 0.06 to 2.8 nm$^{-1}$. The radiation wavelength was $\lambda$ = 0.1488 nm. The intensities were corrected for the detector response and the dark current signal. Distributions of scattered intensity from 2D SAXS images were converted to 1D using the FIT2D software.[41] SAXS curves for samples containing the complex salts $C_{12}$TA(PA$_{100}$-*b*-

PAm$_{93}$), C$_{12}$TA(PA$_{100}$-*b*-PAm$_{366}$), C$_{12}$TA(PA$_{70}$-*b*-PAm$_{422}$) and C$_{12}$TA(PA$_{70}$-*b*-PAm$_{844}$) in water presented a typical region with three broad peak diffractions, ascribed to a cubic *Pm3n* arrangement. Data were fitted to mathematical models using in-house Matlab® routines, for the fitting of the diffraction peaks in terms of wave-vector position $q_i$, amplitude and full-width-half maximum $\Delta q_{FWHM}$ (S5).

## III – Results and Discussion
### III.1 - Light Scattering

Aqueous mixtures of complex salts of C$_{12}$TA(PA$_x$-*b*-PAm$_y$) were first studied by dynamic light scattering measurements. The correlation functions $g^{(2)}(\tau)$ obtained for dilute solutions (*c* = 0.01 and 0.05 wt.%) are shown in Figs. 1a - 1d as a function of the delay time for different scattering angles. The data were treated using the algorithm REPES (GENDIST software)[42] to retrieve the relaxation time distribution functions $\tau A(\tau)$. These distributions are shown in Fig. 2 and indicate the existence of a unique and narrow diffusion mode at all angles.

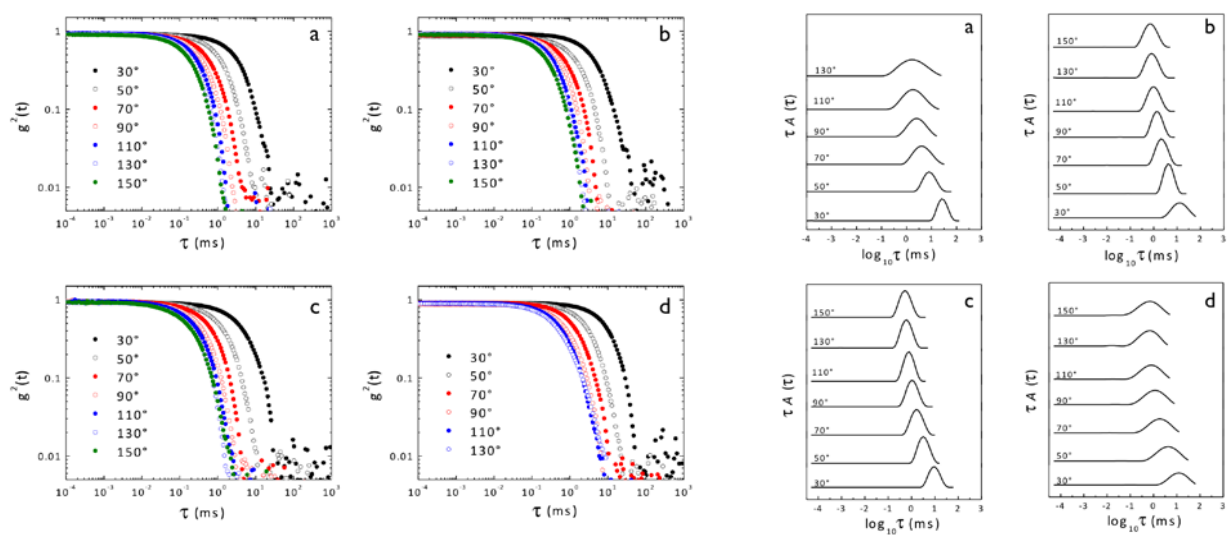

**Figure 1**. *Second autocorrelation function as a function of delay time τ for aqueous mixtures of complex salts: (a) C$_{12}$TA(PA$_{100}$-b-PAm$_{93}$), (b) C$_{12}$TA(PA$_{100}$-b-PAm$_{366}$), (c) C$_{12}$TA(PA$_{70}$-b-PAm$_{422}$) and (d) C$_{12}$TA(PA$_{70}$-b-PAm$_{844}$) at different angles.*
**Figure 2**. *Distribution of relaxation time for aqueous mixtures of complex salts (a) C$_{12}$TA(PA$_{100}$-b-PAm$_{93}$), (b) C$_{12}$TA(PA$_{100}$-b-PAm$_{366}$), (c) C$_{12}$TA(PA$_{70}$-b-PAm$_{422}$) and (d) C$_{12}$TA(PA$_{70}$-b-PAm$_{844}$) at different angles.*

The decay rate $\Gamma(q)$ associated to this diffusion mode is plotted *versus* $q^2$ in Fig. 3 for the 4 solutions. The linear dependence seen in Fig. 3 confirms the diffusive character of the relaxation, and allows to derive the values for the diffusion coefficient $D_0$ and for the hydrodynamic radius $R_H$. For complex salts made from HPA$_{100}$-*b*-PAm$_{93}$, HPA$_{100}$-*b*-PAm$_{366}$, HPA$_{70}$-*b*-PAm$_{422}$ and HPA$_{70}$-*b*-PAm$_{844}$, $R_H$ were 302, 139, 99 and 128 nm, respectively (Table 2). For the symmetric

HPA$_{100}$-b-PAm$_{93}$, the aggregate sizes are large and are associated with strongly scattering solutions.

To get more insight into the morphology of the aggregates, static light scattering was also performed. Fig. 4 shows the variation of the scattering intensity as a function of the scattering angle using the Berry representation, which consists in plotting $\sqrt{\frac{Kc}{R(q,c)}}$ versus $q^2$, where K is the scattering contrast and $R(q,c)$ is the Rayleigh ratio. Previous studies have shown that this representation is more appropriate than the Debye or Zimm representations to describe the scattering by large aggregates (> 100 nm).[37] Fig. 4 shows that $\sqrt{\frac{Kc}{R(q,c)}}$ is a linear increasing functions of $q^2$. From the slope and intercepts, the values for $R_G$ and $M_W^{App}$ were obtained and listed in Table 2. $R_G$-values range from 70 to 250 nm and show the same behavior as $R_H$ as a function of copolymer asymmetry. The ratio between $R_G$ and $R_H$ defines a dimensionless parameter that is associated with the architecture of the aggregates.[43] For $C_{12}TA(PA_x$-b-PAm$_y)$ complex salts, $R_G/R_H$ are comprised between 0.66 to 0.82, in good agreement with the value of a sphere ($\sqrt{3/5}$). The apparent molecular weight of the aggregates obtained from the extrapolation of the Berry plots at $q \rightarrow 0$ was found in the range $9\times10^5$ to $7\times10^8$ g mol$^{-1}$ (Table 2). $M_W^{App}$-values are analyzed in the next section.

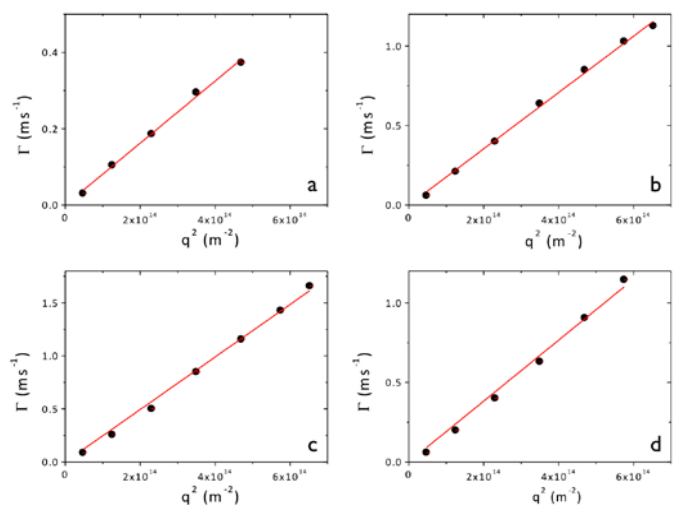

*Figure 3.* Graphs of the decay rate $\Gamma(q)$ in function of $q^2$ for aqueous mixtures of (a) $C_{12}TA(PA_{100}$-b-PAm$_{93})$, (b) $C_{12}TA(PA_{100}$-b-PAm$_{366})$, (c) $C_{12}TA(PA_{70}$-b-PAm$_{422})$ and (d) $C_{12}TA(PA_{70}$-b-PAm$_{844})$.

Colloidal complexes obtained from HPA$_x$-b-PAm$_y$ and $C_{12}$TAB were reported in the literature, however following different pathways. First attempts were made by directly mixing polymer (HPA$_{70}$-b-PAm$_{422}$ and HPA$_{70}$-b-PAm$_{844}$) and surfactant solutions at fixed concentration and mixing ratio.[7,19] More recently, titration experiments were carried out. Titration consists in adding surfactant to polymer, or the reverse, by adding polymer to surfactant in a step-wise manner, exploring thus a wide range of charge ratio.[21] By the direct mixing method, core-shell aggregates with hydrodynamic radii $R_H$ = 30 nm were obtained, whereas titration provided slightly larger values, around 40 and 50 nm depending on the mixing order. In these two cases, the sizes of the aggregates were much smaller than those obtained by the complex salt approach. Similar findings were found for the system $C_{12}TAB$ and $HPA_{70}$-$b$-$PAm_{844}$.[19] In conclusion, light scattering evidences the formation of colloids made from complex salts, their sizes being larger than those obtained via other mixing pathways.

| Complex salt | $D_0$ (m² s⁻¹) | $R_H$ (nm) | $R_G$ (nm) | $M_W^{App}$ (g mol⁻¹) | $R_G/R_H$ | n |
|---|---|---|---|---|---|---|
| $C_{12}TA(PA_{100}$-$b$-$PAm_{93})$ | 8.1×10⁻¹³ | 302 | 247 | 7.2×10⁸ | 0.82 | 37000 |
| $C_{12}TA(PA_{100}$-$b$-$PAm_{366})$ | 1.8×10⁻¹² | 139 | 95 | 8.2×10⁷ | 0.68 | 2700 |
| $C_{12}TA(PA_{70}$-$b$-$PAm_{422})$ | 2.5×10⁻¹² | 99 | 71 | 3.6×10⁷ | 0.72 | 900 |
| $C_{12}TA(PA_{70}$-$b$-$PAm_{844})$ | 1.9×10⁻¹² | 128 | 84 | 9.2×10⁵ | 0.66 | 50* |

**Table 2.** Characteristics of aggregates resulting from dissolution of $C_{12}TA(PA_x$-$b$-$PAm_y)$ complex salt in water. $D_0$ denotes the diffusion coefficient of the aggregates, $R_H$ and $R_G$ the hydrodynamic and gyration radii, $M_W^{App}$ the molecular weight and $n$ the aggregation number expressed in terms of number of surfactant micelles per aggregates. The star (*) indicates that the aggregation number was calculated using the Zimm model for the analysis of the static light scattering data (instead of the Berry diagram for the three other samples).

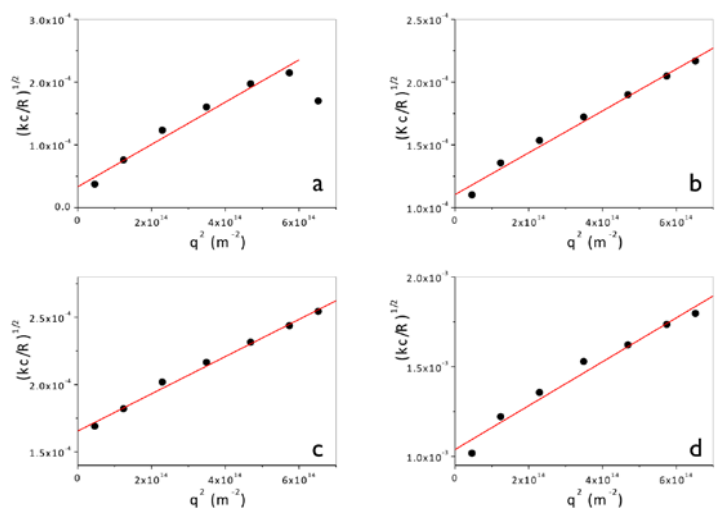

*Figure 4. Graphs of Berry for the aqueous mixtures of (a) $C_{12}TA(PA_{100}$-$b$-$PAm_{93})$, (b) $C_{12}TA(PA_{100}$-$b$-$PAm_{366})$, (c) $C_{12}TA(PA_{70}$-$b$-$PAm_{422})$ and (d) $C_{12}TA(PA_{70}$-$b$-$PAm_{844})$.*

### III.2 - Small-Angle X-ray scattering (SAXS)

Aqueous mixtures of complex salts $C_{12}TA(PA_x\text{-}b\text{-}PAm_y)$ were analyzed by SAXS to assess the internal structure of the aggregates. Figs. 5a-d show the scattering intensities for the four complex salts at concentrations ranging from 5 to 60 wt.% and room temperature ($T = 25\ °C$). In the wave-vector range $q = 1 - 3$ nm$^{-1}$, the intensity exhibits a broad scattering peak, as well as a series of well-defined Braggs reflections. These Braggs reflections attest of the existence of a long-range order. For the system $C_{12}TA(PA_{70}\text{-}b\text{-}PAm_{844})$, the long-range order only shows up at the highest concentration, $c = 60$ wt.%. The relative positions of the diffraction peak in Fig. 5 and noted $q_{h,k,l}$ exhibit the sequence:

$$q_{h,k,l} = \frac{2\pi}{a}\sqrt{h^2 + k^2 + l^2} \qquad (3)$$

where $(h, k, l)$ denote the Miller indices and $a$ the size of the unit cell. Here the sequence of peaks corresponds to crystallographic planes (1,1,0), (2,0,0), (2,1,0), (2,1,1), (2,2,0) etc… and is in agreement with a $Pm3n$ cubic structure.[19,44–46]

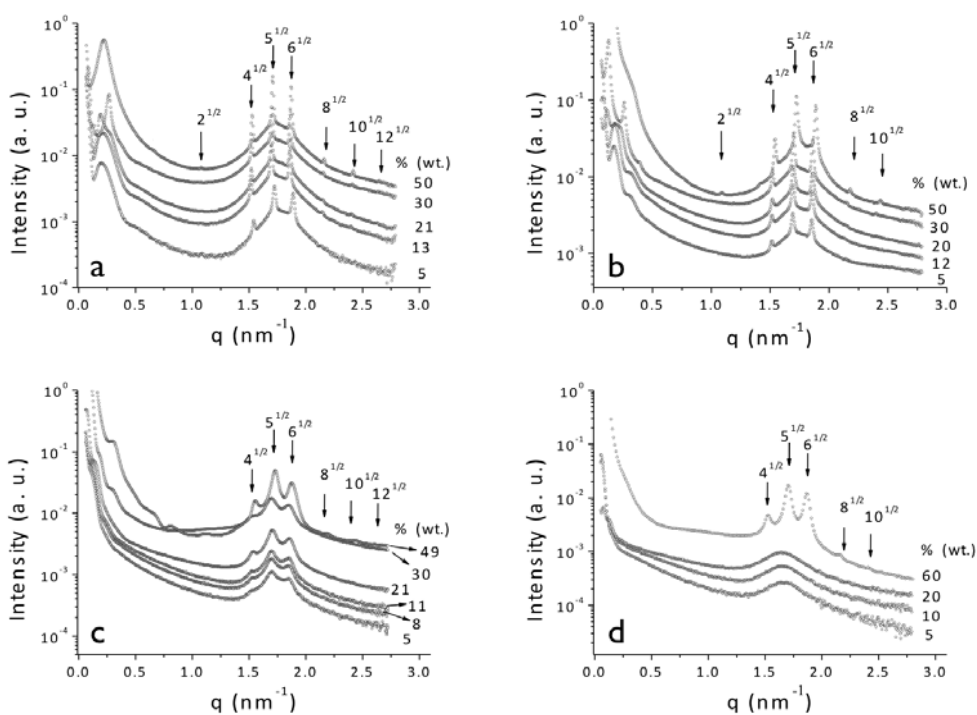

**Figure 5.** *SAXS curves for different concentrations of complex salt in water: (a) $C_{12}TA(PA_{100}\text{-}b\text{-}PAm_{93})$; (b) $C_{12}TA(PA_{100}\text{-}b\text{-}PAm_{366})$; (c) $C_{12}TA(PA_{70}\text{-}b\text{-}PAm_{422})$ and (d) $C_{12}TA(PA_{70}\text{-}b\text{-}PAm_{844})$.*

Features of $Pm3n$ space group are a primitive cell with two micelles of each side of the cube, and a maximum volume fraction of 0.524.[44,46] The lattice parameter $a$ deduced from the $q_{h,k,l}$'s for the different complex salts were found to be independent of the polymer molecular weight and of the asymmetry. Of the order of 8.2 nm, the distance $a$ is in good agreement with those found in the literature for mesophases precipitated from mixtures of $C_{12}TAB$ and poly(acrylic acid),[19,44–46] as well as for $C_{12}TA(PA_x)$ complex salts.[27] Concerning the later system, Svensson and co-workers studied $C_{12}TA(PA_{30})$ and $C_{12}TA(PA_{6000})$ structures prepared according to the same protocols, and found precipitates with a $Pm3n$ symmetry. Moreover, a unit cell size of 8.3 nm was found to be independent on the molecular weight of the polyanion. This finding suggests that the aggregates seen by light scattering have an internal structure which is liquid crystalline and that it is identical to that obtained by precipitation of poly(acrylic acid) and $C_{12}TAB$ or by using their complex salts.[45] The complete list of lattice parameters obtained in this and in further structures is given in Supporting Information (S4).

From analysis of the width of the diffraction peaks $\Delta q_{FWHM}$, the size of the crystallites $\xi$ can be obtained using the Debye-Scherrer equation:[47]

$$\xi = k \frac{2\pi}{\Delta q_{FWHM} - \Delta q_{Res}} \quad (4)$$

where $k = 0.9$[48] and $\Delta q_{Res}$ is a measure of the experimental resolution (here $\Delta q_{Res}/q = 1\%$). The width of the peaks were extracted by fitting the intensity in the range $q = 1 - 3$ nm$^{-1}$ by the sum of four Gaussian functions with adjustable amplitude, position and broadening. One Gaussian aimed to match the broad peak observed in this range, whereas the three others account for the (2,0,0), (2,1,0), (2,1,1) Bragg reflections. Examples of best fit calculations are illustrated in the Supporting Information (S5). For $C_{12}TA(PA_{100}\text{-}b\text{-}PAm_{93})$, $C_{12}TA(PA_{100}\text{-}b\text{-}PAm_{366})$ and $C_{12}TA(PA_{70}\text{-}b\text{-}PAm_{422})$ complex salts at $c = 5$ and 10 wt.%, $\xi$ was found to be 160, 145 and 95 nm respectively. Note that the crystallite sizes $\xi$'s are typically 50% smaller that the hydrodynamic diameters of the aggregates found by light scattering (Table 2). This finding suggests that they represent an order of magnitude of the crystalline core. Quantitative determinations in terms of core and shell distances remains however delicate here, and should be supplemented by additional experiments, e.g. small-angle scattering at lower wave-vector.

From the values of the molecular weights $M_W^{App}$ listed in Table 2, a description of the microstructure in terms of aggregation number (*i.e.* in terms of numbers of micelles and polymers per aggregate) is achievable. It is based on the assumption that the inner structure of the aggregates is liquid-crystalline, with micelles arranged on a cubic $Pm3n$ lattice. It assumes moreover that the complexes are formed at the charge stoichiometry (*i.e.* there are as many surfactants as carboxylate monomers), and that $C_{12}TAB$ micelles have a fixed aggregation number (taken to be 53 here),[20] namely that of $C_{12}TAB$ micelles at the cmc. The number of micelles $n$ per aggregate is presented Table 2. $n$-values range from 50 to $3.7 \times 10^4$ and follow the same tendency as those observed for the hydrodynamic and gyration radii: $n$ increases with the asymmetry of the diblock. Note that the value of $n = 50$ found for $C_{12}TA(PA_{70}\text{-}b\text{-}PAm_{844})$ was calculated using the Zimm approximation, instead of that of Berry for the adjustment of the light scattering data. This value is in better agreement with the hydrodynamic and gyration sizes. In Table 2, $C_{12}TA(PA_{100}\text{-}b\text{-}PAm_{366})$ and $C_{12}TA(PA_{70}\text{-}b\text{-}Pam_{844})$ complexes are of comparable sizes, however their molecular weight differs significantly. This effect is due to the fact that aggregates with cubic ordered cores are denser than disordered ones, and also that for $C_{12}TA(PA_{70}\text{-}b\text{-}Pam_{844})$ the shell thickness is larger.

As already mentioned, colloidal complexes made with $HPA_x\text{-}b\text{-}PAm_y$ and $C_{12}TAB$ were reported in the literature using alternative protocols. For $C_{12}TAB$ and $HPA_{70}\text{-}b\text{-}PAm_{422}$ for instance, small-angle neutron scattering allowed to estimate the aggregation numbers (number of micelles and of polymers per colloid) in the core. The SANS study was consistent with aggregation numbers of the order of 100 – 200. These values are much lower than those found for the $C_{12}TA($

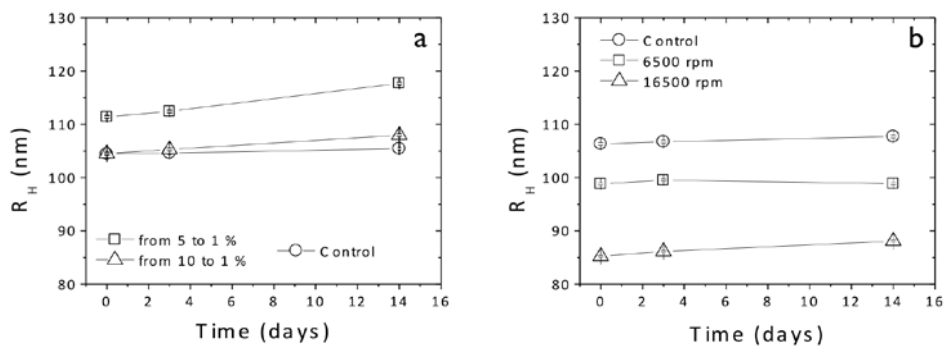

**Figure 6.** *Hydrodynamic radius $R_H$ as a function of the time for $C_{12}TA(PA_{70}\text{-}b\text{-}PAm_{422})$-aggregates formed following the complex salt approach. In this assay, different mixing conditions were used: a) variation of the initial concentration; b) variation of the mixing velocity. Samples were studied at c = 1 wt.% and T = 25 °C.*

### III.3 – Role of the mixing pathway, stability of the cubic structures

Figs. 6 display the hydrodynamic radius $R_H$ as a function of the time for $C_{12}TA(PA_{70}\text{-}b\text{-}PAm_{422})$ complex salt solutions prepared using different mixing pathways. In Fig. 6a, polymer/surfactant aggregates were prepared at different concentrations, from $c$ = 1 to 5 and 10 wt.%. The samples at $c$ = 5 and 10 wt.% were diluted down to $c$ = 1 wt.% for the DLS runs. In Fig. 6b, the samples were prepared using a mechanical disperser (Turrax ®) operating at different mixing velocities between 0 and 16500 rpm. In both experiments, it is shown that the hydrodynamic radius depends slightly on the mixing pathway, $R_H$ varying typically by 10% by changing the initial concentration or the stirring speed. These results support those of earlier reports,[7] and suggest that at the time scale of laboratory experiments, the aggregates prepared from complex salts are out-of-equilibrium colloids. Their final structure depends on the pathway, however this dependence is here weaker.[49–51]

With direct mixing, the $C_{12}TAB$ micelles located in the core were found to be disordered.[7,20] It is hence interesting to study the stability and the structural changes of the liquid crystalline cores upon the addition of various additives, including electrolytes or polyelectrolytes. Fig. 7a – 7c show the SAXS intensity in the 1 – 3 nm$^{-1}$ $q$-range for $C_{12}TA(PA_{100}\text{-}b\text{-}PAm_{366})$ complex salts aqueous solutions prepared with increasing amounts of a) sodium bromide (NaBr), b) trimethylammonium bromide (($CH_3$)$_4$NBr) and (c) poly(acrylic acid) ($HPA_{100}$). For the salts, the electrolytes concentration is expressed in terms of concentration (0 - 0.3 mol L$^{-1}$) whereas for $HPA_{100}$ the amount of added polymers is described in terms of chain ratio $n_{Homo}$ / $n_{copo}$ where $n_{Homo}$ and $n_{copo}$ the numbers of $HPA_{100}$ and $PA_{100}\text{-}b\text{-}PAm_{366}$ present in the dispersion, respectively. As for the electrolytes, the complex salt liquid crystalline structures persist up to a concentration of about 0.26 – 0.28 mol L$^{-1}$.[52] Above, the Bragg reflection vanished into the broad scattering pattern found at 1.6 nm$^{-1}$. Similarly, with the addition of homopolyelectrolytes, the long-range order inside the cores was lost above a chain ratio of 0.6. Interestingly, the loss of crystallinity is associated to a shift of the scattering peak, indicating the increase or collapse of the structures upon addition of electrolytes or of polyelectrolytes, respectively. Similar results were observed on surfactant/polymer complex salts.[27,36] The results indicate that the screening of the electrostatics by addition of electrolytes, or the overcharging of polyions species, here poly(acrylic acid) are able to disrupt the liquid-crystalline state of the complex salts.

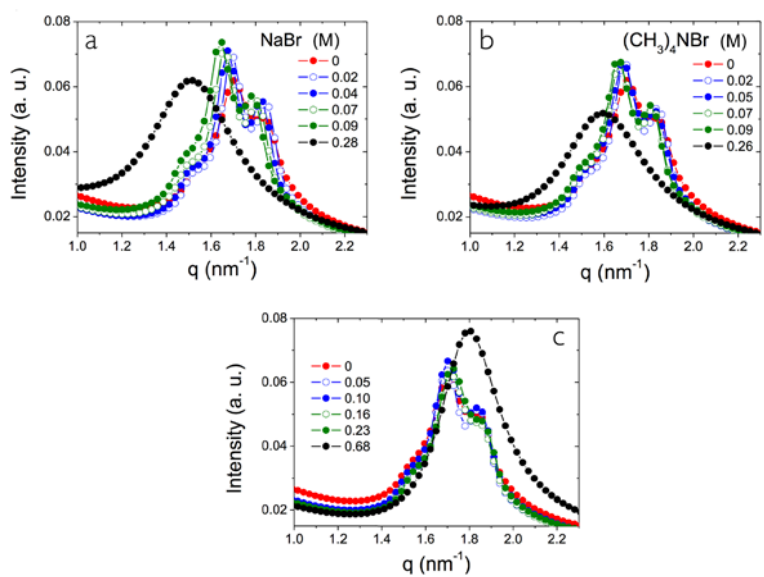

*Figure 7.* SAXS curves for the complex salt $C_{12}TA(PA_{100}$-b-$PAm_{366})$ in water with addition of different quantities of (a) NaBr; (b) $(CH_3)_4NBr$ and (c) poly(acrylic acid).

## IV – Concluding Remarks

Complex salts are single component systems that were developed to investigate the binary phase diagrams of complex mixtures. In the present work, we used complex salts made of charged surfactants and block copolymers to design dispersions of stable hierarchical colloids. The effect of asymmetry (i.e. the ratio between the degrees of polymerization of the charged and neutral blocks) of the copolymer on the colloid microstructure was studied in much detail. For symmetric diblocks, the mixed colloids exhibit a *Pm*3*n* cubic liquid crystalline structure characterized by densely packed surfactant micelles. These hierarchical structures could be useful for the synthesis of porous colloids with controlled pore sizes. The cubic structure and lattice cell dimension were found in excellent agreement with those obtained under the same conditions with homopolyelectrolytes.[27] This finding suggests that the driving force for the precipitated phases and the core-shell colloids is similar. With diblocks however, the growth of the liquid-crystalline phase is arrested at the size of colloids. For low asymmetry ratios, typically below 0.1, self-assembly still occurs, but the internal structure is now disordered (Fig. 5d).

Combining light and X-ray scattering experiments, the microstructure of the surfactant/polymer aggregates was resolved. Both the number of micelles per aggregate and the sizes of the aggregates exhibit similar features: they decrease with increasing diblock asymmetry. With symmetric copolymers, sub-micron colloids were obtained, the aggregation number being several ten thousands. Asymmetric copolymers by contrast self-assemble spontaneously into much smaller objects of the order of 100 nm.[19] It is interesting to notice that an increase in the length of the PAm block (and also of the related asymmetry ratio) leads to a decrease in the aggregates hydrodynamic radii. This may be related to the requirement of a higher curvature object to accommodate the larger PAm block at its interface, somehow related to the use of the critical packing parameter to explain geometry and curvature of surfactant aggregates. The previous assumption is valid for smaller objects and is probably not applicable to the larger aggregates, such as those formed by $HPA_{100}$-$b$-$PAm_{93}$. Similar proposal was recently put forward by Van der Kooij to account for a sphere-to-rod transition on complexes of oppositely charged polyelectrolytes.[53]

At this point, it is important to compare the present results with those of the literature. The very same batch of $HPA_{70}$-$b$-$PAm_{422}$ block copolymer was investigated together with $C_{12}TAB$ using the direct mixing and titration methods.[19,21] With these protocols, mixed surfactant/polymer colloids were found to form spontaneously. Differences with the complex salts exist however and they are significant: *i)* The size and aggregation numbers are much smaller with direct mixing and titration, typically 10 times smaller in terms of aggregation number and 3 times in terms of size. *ii)* In the core of the aggregates, the micelles are disordered, and do not exhibit the *Pm3n* cubic phase characteristic of the complex salts. *iii)* The size and morphology are strongly sensitive to the way the surfactant and polymer were mixed. This comparison deserves an additional comment. With direct mixing, cores of the order of 10 – 20 nm were observed. Such a structure is consistent with the spatial separation of the different blocks, the poly(acrylic acid) being included in the cores with the surfactant micelles and the poly(acrylamide) forming the corona. With the complex salt approach however, the situation is different: the aggregates are in the range 200 – 500 nm. Such sizes are incompatible with a complete separation of the blocks and hence the neutral poly(acrylamide) blocks, or parts of these blocks need to be located within the cubic micro-phase. Moreover, the aggregates display an internal structure that is similar to the equilibrium structure, but remain in the form of a kinetically stable dispersion.

In conclusion, the above results suggest that the complex salt methodology presents many advantages as compared to other protocols. It is possible to prepare new core-shell aggregates at the charge stoichiometric and displaying liquid-crystalline interiors. Finally, it was also found that complex salt aggregates are less sensitive to the preparation conditions as compared to the other protocols.

## ASSOCIATED CONTENT

**Supporting Information**

Preparation and characterization of polymers. Lattice parameters of cubic *Pm3n* phases and analysis of the Bragg diffraction patterns. This material is available free of charge via the Internet at http://pubs.acs.org.

## AUTHOR INFORMATION

# Langmuir

pubs.acs.org/Langmuir

Article

**Corresponding Authors**
* E-mail: [1] wloh@iqm.unicamp.br and [2] jean-francois.berret@univ-paris-diderot.fr
**Author Contributions** The manuscript was written through contributions of all authors. All authors have given approval to the final version of the manuscript.

ACKNOWLEDGMENT
This work was supported by a bilateral cooperation between CNRS and FAPESP (Proc. No. 2010/52411-6). Authors gratefully acknowledge the Brazilian Synchrotron Laboratory (LNLS) for allocation of SAXS beamtime. LV thanks the Brazilian Agency Capes for a PhD fellowship, and WL thanks CNPq for a senior researcher grant.